\documentclass[submitted,copyright,sharealike,noncommercial]{eptcs}
\providecommand{\event}{UITP 2016} 
\usepackage{breakurl}              
\usepackage{float}
\newfloat{listing}{tbhp}{lst}
\floatname{listing}{Listing}
\usepackage{xspace}
\usepackage{graphicx}
\usepackage{subcaption}
\usepackage{color}
\usepackage{tikz}
\usetikzlibrary{shadows,calc}
\definecolor{lightgray}{rgb}{0.95, 0.95, 0.95}
\definecolor{darkgray}{rgb}{0.4, 0.4, 0.4}
\definecolor{editorGray}{rgb}{0.95, 0.95, 0.95}
\definecolor{editorOcher}{rgb}{1, 0.5, 0} 
\definecolor{editorGreen}{rgb}{0, 0.5, 0} 
\definecolor{orange}{rgb}{1,0.45,0.13}
\definecolor{olive}{rgb}{0.17,0.59,0.20}
\definecolor{brown}{rgb}{0.69,0.31,0.31}
\definecolor{purple}{rgb}{0.38,0.18,0.81}
\definecolor{lightblue}{rgb}{0.1,0.57,0.7}
\definecolor{lightred}{rgb}{1,0.4,0.5}

\usepackage{listings}
\lstdefinelanguage{CSS}{
  keywords={color,background-image:,margin,padding,font,weight,display,position,top,left,right,bottom,list,style,border,size,white,space,min,width, transition:, transform:, transition-property, transition-duration, transition-timing-function},
  sensitive=true,
  morecomment=[l]{//},
  morecomment=[s]{/*}{*/},
  morestring=[b]',
  morestring=[b]",
  alsoletter={:},
  alsodigit={-}
}

\lstdefinelanguage{JavaScript}{
  morekeywords={typeof, new, true, false, catch, function, return, null, catch, switch, var, if, in, while, do, else, case, break, val, object, module, where},
  morecomment=[s]{/*}{*/},
  morecomment=[l]//,
  morestring=[b]",
  morestring=[b]'
}

\lstdefinelanguage{HTML5}{
  language=html,
  sensitive=true,
  alsoletter={<>=-},
  morecomment=[s]{<!-}{-->},
  tag=[s],
  otherkeywords={
  >,
	<!DOCTYPE,
  </html, <html, <head, <title, </title, <style, </style, <link, </head, <meta, />,
	</body, <body,
	</div, <div, </div>,
  </span, <span, </span>,
  </section, <section, </section>,
	</p, <p, </p>,
  </code, <code, </code>,
  </h1, <h1, </h1>,</h2, <h2, </h2>,</h3, <h3, </h3>,
	</script, <script,
  <canvas, /canvas>, <svg, <rect, <animateTransform, </rect>, </svg>, <video, <source, <iframe, </iframe>, </video>, <image, </image>, <header, </header, <article, </article
  },
  ndkeywords={
  =,
  charset=, src=, id=, width=, height=, style=, type=, rel=, href=, class=, id=
  fill=, attributeName=, begin=, dur=, from=, to=, poster=, controls=, x=, y=, repeatCount=, xlink:href=,
  margin:, padding:, background-image:, border:, top:, left:, position:, width:, height:, margin-top:, margin-bottom:, font-size:, line-height:,
  transform:, -moz-transform:, -webkit-transform:,
  animation:, -webkit-animation:,
  transition:,  transition-duration:, transition-property:, transition-timing-function:,
  }
}

\lstdefinestyle{htmlcssjs} {%
  basicstyle={\footnotesize\ttfamily},
  identifierstyle=\color{black},
  keywordstyle=\color{blue}\bfseries,
  ndkeywordstyle=\color{editorGreen}\bfseries,
  stringstyle=\color{editorOcher}\ttfamily,
  commentstyle=\color{brown}\ttfamily,
  language=HTML5,
  alsolanguage=JavaScript,
  morecomment=[s]{\{-}{-\}},
  alsodigit={.:;},
  tabsize=2,
  showtabs=false,
  showspaces=false,
  showstringspaces=false,
  extendedchars=true,
  breaklines=true,
  columns=fullflexible,
  literate=%
  {Ö}{{\"O}}1
  {Ä}{{\"A}}1
  {Ü}{{\"U}}1
  {ß}{{\ss}}1
  {ü}{{\"u}}1
  {ä}{{\"a}}1
  {ö}{{\"o}}1
}

\lstdefinestyle{isabelle} {%
  basicstyle={\footnotesize\sffamily},
  tabsize=2,
  showtabs=false,
  showspaces=false,
  showstringspaces=false,
  extendedchars=false,
  breaklines=true,
  columns=fullflexible
}

\def\shadowshift{1pt,-1pt}
\def\shadowradius{4pt}

\colorlet{innercolor}{black!60}
\colorlet{outercolor}{gray!05}

\newcommand\drawshadow[1]{
    \begin{pgfonlayer}{shadow}
        \shade[outercolor,inner color=innercolor,outer color=outercolor] ($(#1.south west)+(\shadowshift)+(\shadowradius/2,\shadowradius/2)$) circle (\shadowradius);
        \shade[outercolor,inner color=innercolor,outer color=outercolor] ($(#1.north west)+(\shadowshift)+(\shadowradius/2,-\shadowradius/2)$) circle (\shadowradius);
        \shade[outercolor,inner color=innercolor,outer color=outercolor] ($(#1.south east)+(\shadowshift)+(-\shadowradius/2,\shadowradius/2)$) circle (\shadowradius);
        \shade[outercolor,inner color=innercolor,outer color=outercolor] ($(#1.north east)+(\shadowshift)+(-\shadowradius/2,-\shadowradius/2)$) circle (\shadowradius);
        \shade[top color=innercolor,bottom color=outercolor] ($(#1.south west)+(\shadowshift)+(\shadowradius/2,-\shadowradius/2)$) rectangle ($(#1.south east)+(\shadowshift)+(-\shadowradius/2,\shadowradius/2)$);
        \shade[left color=innercolor,right color=outercolor] ($(#1.south east)+(\shadowshift)+(-\shadowradius/2,\shadowradius/2)$) rectangle ($(#1.north east)+(\shadowshift)+(\shadowradius/2,-\shadowradius/2)$);
        \shade[bottom color=innercolor,top color=outercolor] ($(#1.north west)+(\shadowshift)+(\shadowradius/2,-\shadowradius/2)$) rectangle ($(#1.north east)+(\shadowshift)+(-\shadowradius/2,\shadowradius/2)$);
        \shade[outercolor,right color=innercolor,left color=outercolor] ($(#1.south west)+(\shadowshift)+(-\shadowradius/2,\shadowradius/2)$) rectangle ($(#1.north west)+(\shadowshift)+(\shadowradius/2,-\shadowradius/2)$);
        \filldraw ($(#1.south west)+(\shadowshift)+(\shadowradius/2,\shadowradius/2)$) rectangle ($(#1.north east)+(\shadowshift)-(\shadowradius/2,\shadowradius/2)$);
    \end{pgfonlayer}
}

\pgfdeclarelayer{shadow}
\pgfsetlayers{shadow,main}

\newsavebox\mybox
\newlength\mylen

\newcommand\shadowimage[2][]{%
\setbox0=\hbox{\includegraphics[#1]{#2}}
\begin{tikzpicture}
\node[anchor=south west,inner sep=0] (image) at (0,0) {\includegraphics[#1]{#2}};
\drawshadow{image}
\end{tikzpicture}}

\title{Interactive Proof Presentations with \textit{Cobra}}
\author{Martin Ring
\institute{DFKI\\ Bremen, Germany}
\email{martin.ring@dfki.de}
\and
Christoph L\"uth
\institute{DFKI and Universit\"at Bremen\\ Bremen, Germany}
\email{christoph.lueth@dfki.de}
}
\def\titlerunning{Interactive Proof Presentations with \textit{Cobra}}
\def\authorrunning{M. Ring \& C. L\"uth}

\newcommand{\eg}{\textit{e.g.}\xspace}

\newcommand{\ie}{\textit{i.e.}\xspace}

\begin{document}

\maketitle

\begin{abstract}
  We present \textit{Cobra}, a modern proof presentation
  framework, leveraging cutting-edge presentation technology together with
  a state of the art interactive theorem prover to present formalized
  mathematics as active documents. \textit{Cobra} provides both an easy way
  to present proofs and a novel approach to auditorium interaction. The
  presentation is checked live by the theorem prover, and moreover
  allows for live changes both by the presenter and the audience.
\end{abstract}

\section{Introduction}

Presenting formalized mathematical proofs is a challenge by
itself. Formalized proofs, by their very nature, tend to be lengthy, as
every side condition and assumption has to be tackled, and often
the amount of proof text devoted to the main argument is small in relation
to those necessary but tedious side issues. Further, the actual proof
usually contains technical information such as setting up the syntactic
machinery and automatic proof assistance of the prover, which is not
necessary for the comprehension of the mathematical argument, but crucial
if the audience wants to rerun the proof in the actual prover.
Thus, when presenting formalized proofs to students, fellow researchers, or
other audiences, we only present a \emph{view} on the actual proof
text. This view only contains excerpts of the original proof, and is
constructed manually; it is
left to the punctiliousness of the presenter to not introduce errors. The
correlation between the presentation and the actual proof text is left to
the critical audience to verify, as the resulting view cannot be checked by
the prover anymore.
Further, when made with traditional presentation aids (such as PowerPoint,
\LaTeX, or writing on plain old blackboards) these presentations are not
interactive --- we cannot easily change them during the presentation, \eg
to demonstrate why a particular approach will not prove the desired
goal. However, a proof made with an \emph{interactive} theorem prover is an
\emph{active document} and should be treated, and presented, as such. The
stop-gap measure often used up to now has been to switch back and forth
between the presentation and the actual running prover, but the resulting
change of focus makes it hard to follow the proof, and as the whole proof
text is shown, the audience is potentially overwhelmed with technical
details. Another fix is to animate the presentation manually by means of
PDF overlays or PowerPoint animations, but this is inflexible, error-prone
and cumbersome.

Fortunately, the advances of modern web technologies have opened up new ways of
presentation, which allow us to treat a proof as an active document
rather than an inanimate piece of PDF. Tools such as \textit{reveal.js}
implement presentations as active documents, and thus it seems logical to
leverage this technology for interactive provers.
In this paper, we present \textit{Cobra}, an integrated presentation
environment for interactive proofs and code. \textit{Cobra} allows us to
declaratively define interactive slides containing Isabelle theories (or
snippets from a theory), \LaTeX{}-style formulae and program
code. The intriguing aspect about these slides is that they can not only be
presented with annotated semantic information provided by the underlying
prover (or compiler in the case of code), but also the content can be
altered (or completed), resulting in updated semantic annotations. This
way, the presenter can develop the proof in front of, and even in
interaction with, the audience, instead of following a strict predetermined
path.

The paper is structured as follows: Section~\ref{sec:ux} describes how to
work with \textit{Cobra} from a users perspective. Section~\ref{sec:int}
sketches the architecture of the project, and Section~\ref{sec:rel-work}
briefly discusses related work. We conclude in Section~\ref{sec:concl} by
describing proposed use cases and providing an outlook onto future work.

\section{Presenting with \textit{Cobra}}
\label{sec:ux}

Creating slides is one of the most time consuming tasks in the preparation
of a lecture or a talk. Thus it should involve no additional overhead, to
allow the speaker to focus on content rather than technical
details. \textit{Cobra} aims to be even simpler than presenting with the
\LaTeX{} beamer class but a lot more powerful.
It comes as a self-contained command line tool\footnote{%
  The binary can be obtained from \url{http://www.flatmap.net/cobra} for
  all major operating systems, or alternatively as source code from
  \url{https://github.com/flatmap/cobra/}.}, %
and supports Isabelle/HOL, Scala and Haskell as well as \LaTeX-style
formulae out of the box, and furthermore can be extended to suit individual
needs.
The only strong prerequisite is an active installation of Java 8 or above.
To check whether the system setup supports
interactive presentation of a specific language, there is the
\texttt{cobra configure} command which accepts a language identifier as an argument (i.e. \texttt{isabelle}, \texttt{scala}, \texttt{haskell}), which
analyses whether the optional prerequisites for the selected language are met
and provides further advice.

\paragraph{To create an empty \textit{Cobra} presentation} just run
\texttt{cobra new <name>} on the command line. This creates a
directory \texttt{<name>} containing two files: \texttt{cobra.conf} and
\texttt{slides.html}. The former contains meta information and environment
configurations, while the latter contains the actual content of the
presentation. The user may add other files to the directory, which will
then be served by \textit{Cobra}. This allows to include graphics, videos,
custom style sheets and other arbitrary files, which can then be included
in the \texttt{slides.html} file.  During presentation, \textit{Cobra} will
run an embedded web server, and serve the slides as HTML pages (see Section~\ref{sec:interactive-presentations}).

\subsection{Configuration}

The \texttt{cobra.conf} file is a HOCON (\emph{Human-Optimised Config
  Object Notation}) \cite{HOCON} configuration file. The file \texttt{reference.conf} is
included in the distribution which contains all available settings together
with default values and short descriptions.  There is no need to change
\texttt{cobra.conf} for most presentations.  However, among others, the
customisations shown in Table~\ref{tab:settings} are available. Other
available settings include the configuration of the underlying presentation
framework \textit{reveal.js}, the math engine \textit{MathJax} as well as
Isabelle. All settings provide reasonable default values (``convention over
configuration'').

\begin{table}[h]
\begin{tabular}{ll}
\texttt{title} & display title of the presentation\\
\texttt{language} & main language of the presentation\\
\texttt{theme.slides} & main style sheet that should be used to render the slides\\
\texttt{theme.code} & main style sheet that should be used to render code snippets\\
\texttt{binding.interface} & network interface to bind the server on (default \texttt{localhost})\\
\texttt{binding.port} & port under which the server will be available (default \texttt{8080})\\
\texttt{reveal.transition} & the default transition between slides (e.g. \texttt{slide}, \texttt{fade}, \texttt{none})\\
\texttt{env.isabelle\_home} & environment variable to be picked up by language servers
\end{tabular}
\caption{Some of the settings available through \texttt{cobra.conf}}
\label{tab:settings}
\end{table}

Changing a setting has immediate effect without reloading the application (except for the network interface which can not be switched while cobra is running), such that the user can always observe the effects of the current configuration in the browser.

\subsection{Adding slides}

The presentation facilities of \textit{Cobra} are based on the
\textit{reveal.js} framework \cite{revealjs}.  The presentation resides in
one HTML file (called \texttt{slides.html}).  However, this file has no top
\texttt{html} element (and is thus not valid HTML); instead, the slides are
represented by top-level \texttt{section} elements and behave as
\textit{reveal.js} slides with the exception of \texttt{code} elements,
which are described in Section~\ref{sec:code}. All the boilerplate needed
to make this a valid HTML document, and add the necessary JavaScript and
style sheets is automatically generated by \textit{Cobra}.

\textit{reveal.js} has a two dimensional slide layout: It is possible to
nest \texttt{section} elements by one level, which results in vertically
arranged slides.  This can be used to group slides together. The content of the
\texttt{section} elements can be arbitrary HTML, allowing for rich
presentations. However, typically a small subset will suffice. Particularly
relevant are header elements (\texttt{h1}, \texttt{h2}, \ldots) for
the title of a slide; \texttt{img} elements to include vector or raster
graphics; and unordered and ordered list elements (\texttt{ul}, \texttt{ol})
for itemisations or enumerations respectively. \textit{reveal.js} supports
so-called fragments, which allow the user to
unhide or emphasise certain parts of the slide. Fragments are added through
class attributes. Listing \ref{lst:slides} contains a small example with
five slides.

\subsection{Integrating \LaTeX-style formulae}

Text surrounded with dollar signs (i.e. \texttt{\$a \textbackslash rightarrow b\$}) is
interpreted as \LaTeX\ and rendered by \textit{MathJax}, a JavaScript library
that can render MathML and a large subset of \TeX / \LaTeX\ and is compatible
with all modern browsers \cite{cervone2012mathjax}. The MathJax configuration
can be altered through the \texttt{cobra.conf} file.

\subsection{Including proofs and code}
\label{sec:code}
As mentioned above, \texttt{code} elements are treated specially by
\textit{Cobra}.  By default, every \texttt{code} element creates an
editable code snippet. By adding certain class attributes to the element,
its behaviour can be altered, e.g. the language mode can be defined by adding
either \texttt{scala}, \texttt{haskell} or \texttt{isabelle} as a
class. When such a class attribute is added, the rich semantic assistance
engine is automatically invoked to provide information about the snippet,
that will be visualised during the presentation.

The simplest way to include code is to add it as content of a \texttt{code}
element. However, it is often desirable to include only certain snippets
of a larger code example. For such situations, \textit{Cobra} allows us to
include hidden \texttt{code} elements and then reference snippets. Snippets
are marked by special comments within the target language, which is more
robust than referencing lines. Snippets may overlap.  When dealing with
large code examples it is also possible to include just a reference to an
external source file. In this case, the language does not have to be
specified as it can be derived from the file extension.

\paragraph{Behavioural Classes} can be added to a \texttt{code} element to
alter its presentation behaviour: \texttt{states} enables inline proof
states (for Isabelle only),
\texttt{state-fragments} reveals one proof after another during the
presentation, \texttt{no-infos} hides info messages during the
presentation, and
\texttt{no-warnings} hides warning messages during the presentation.

\paragraph{Code Fragments} are parts of code that are revealed or altered interactively during the presentation. There is a special comment syntax which preserves the semantics of the original source file. The variants of a section of source code are enclosed in parentheses and separated by a pipe symbol. The parentheses are then enclosed in block comments, including either the left or right variant. e.g. \lstinline[{style=htmlcssjs}]{val x = /*(*/???/*|3 * 7)*/} or \lstinline[{style=htmlcssjs}]{val x = /*(???|*/3 * 7/*)*/}, which will both result in the same sequence during the presentation but obviously produce different meanings in the source file.
When no alternative is provided (as in \textbf{lemma} $x$: $A \Rightarrow$ \texttt{(*(*)}$A$\texttt{(*)*)}) the code fragment
is selected during the presentation and semantic information is annotated if available.

\begin{figure}
\begin{subfigure}[b]{.5\textwidth}
\begin{lstlisting}[style=htmlcssjs]
<code class="hidden" src="src/Seq.thy">
</code>
<section>
  <h2>A Short Demo</h2>
  Sequences and their concatenation
  <code src="#def-seq-conc">
  </code>
</section>
<section>
  <h2>A Short Lemma</h2>
  <code src="#reverse-conc" class="states">
  </code>
</section>
<section>
  <h2>A Short Proof</h2>
  <code src="#reverse-reverse" class="states">
  </code>
</section>
<section>
  <h2>A Short Haskell Demo</h2>
  <code class="haskell">
    module Example where
    fibs = {-(-}undefined{-|0 : 1 : zipWith (+) fibs (tail fibs))-}
  </code>
</section>
<section>
  <h2>A Short Scala Demo</h2>
  <code class="scala">
    object Example {
      val x = /*(???|*/3 * 7/*)*/
    }
  </code>
</section>

\end{lstlisting}
\subcaption{\texttt{slides.html}\label{lst:slides}}
\end{subfigure}
\begin{subfigure}[b]{.5\textwidth}
\renewcommand{\baselinestretch}{0.92}\normalsize
\begin{footnotesize}
\begin{tabbing}
\textbf{theory} Seq\\
\textbf{imports} Main\\
\textbf{begin}\\
\\
\texttt{(** begin \#def-seq-conc *)}\\
\textbf{datatype} $'\alpha$ seq $=$ Empty $|$ Seq $'\alpha$ $'\alpha$ seq\\
\\
\textbf{fun} conc :: $'\alpha$ seq $\Rightarrow$ $'\alpha$ seq $\Rightarrow$ $'\alpha$ seq\\
\textbf{where}\\
\quad \= conc Empty $ys$ \hspace{1.4em}\=$=$ $ys$\\
$|$ \> conc (Seq $x$ $xs$) $ys$ \>$=$ Seq $x$ (conc $xs$ $ys$)\\
\texttt{(** end \#def-seq-conc *)}\\
\\\kill
\textbf{fun} reverse :: $'\alpha$ seq $\Rightarrow$ $'\alpha$ seq\\
\textbf{where}\\
    \> reverse Empty \hspace{1.4em}\=$=$ Empty\\
$|$ \> reverse (Seq $x$ $xs$) \>$=$ conc (reverse $xs$) (Seq $x$ Empty)\\
\\
\textbf{lemma} conc\_empty: conc $xs$ Empty $=$ $xs$\\
  \> \textbf{by} (induct $xs$, simp\_all)\\
\\
\textbf{lemma} conc\_assoc: conc (conc $xs$ $ys$) zs $=$ conc $xs$ (conc $ys$ zs)\\
  \> \textbf{by} (induct $xs$, simp\_all)\\
\\
\texttt{(** begin \#reverse-conc *)}\\
\textbf{lemma} reverse\_conc: \\
\quad \quad reverse (conc $xs$ $ys$) $=$ conc (reverse $ys$) (reverse $xs$)\\
  \> \textbf{apply} (induct $xs$)\\
  \> \textbf{apply} (simp\_all add: conc\_empty conc\_assoc)\\
  \> \textbf{done}\\
\texttt{(** end \#reverse-conc *)}\\
\\
\texttt{(** begin \#reverse-reverse *)}\\
\textbf{lemma} reverse\_reverse: reverse (reverse $xs$) $=$ $xs$\\
  \> \textbf{oops}\\
\texttt{(** end \#reverse-reverse *)}\\
\\
\textbf{end}
\end{tabbing}
\end{footnotesize}
\subcaption{\texttt{src/Seq.thy}\label{lst:seq}}
\end{subfigure}
\caption{Simple example with five slides and external Isabelle source.}
\renewcommand{\baselinestretch}{1.00}\normalsize
\end{figure}

\begin{figure}[ht]
\centering
\shadowimage[width=.9\linewidth]{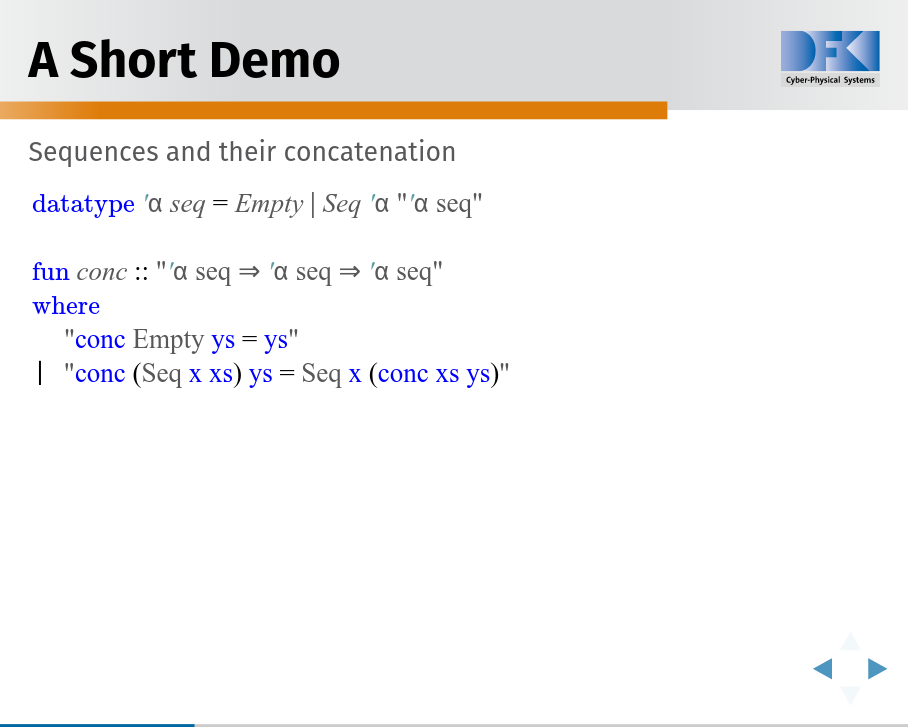}
\caption{The rendered result of Listing~\ref{lst:slides}\label{fig:slides}. The syntax highlighting for the inner syntax (i.e. the double quoted parts) is provided by the prover.}
\end{figure}

\subsection{Interactive presentations}
\label{sec:interactive-presentations}

\begin{figure}[h]
\centering
\shadowimage[width=.9\linewidth]{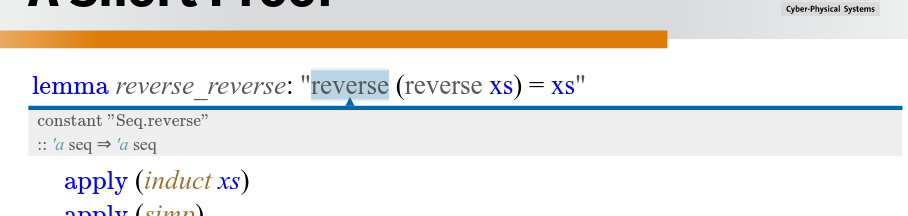}
\caption{A hover tool-tip displays information about ``reverse''\label{fig:hover}}
\end{figure}

\begin{figure}[h]
\centering
\shadowimage[width=.9\linewidth]{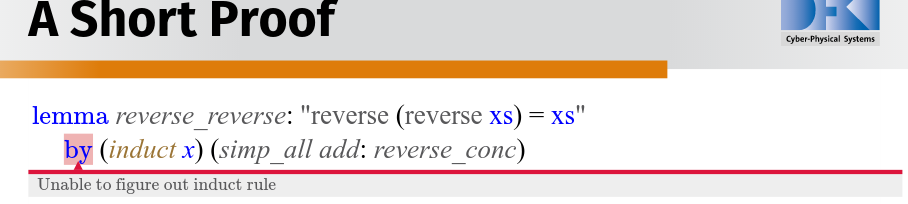}
\caption{An error is displayed\label{fig:error}}
\end{figure}

\begin{figure}[h]
\centering
\shadowimage[width=.9\linewidth]{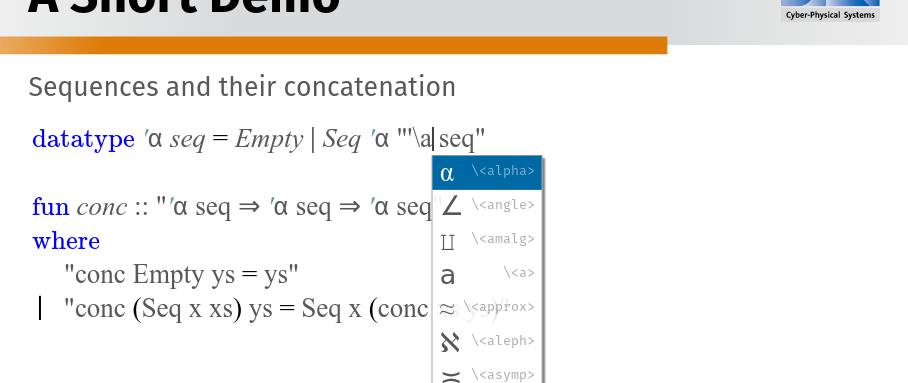}
\caption{Editing the proof during the presentation\label{fig:edit}}
\end{figure}

\begin{figure}[h]
\centering
\shadowimage[width=.9\linewidth]{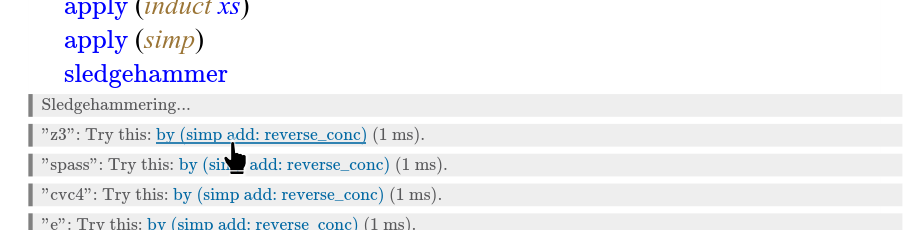}
\caption{Using a result from an asynchronous task\label{fig:sendback}}
\end{figure}

Use \texttt{cobra <dir>} on the command line to
start a presentation server within the specified directory. After a short while
\textit{Cobra} will be initialised.
The presentation can now be opened with any modern web browser.
The user will see the first slide. It is possible to navigate through the slides
with the keyboard or a presenter.
\textit{reveal.js} provides controls similar to those of PowerPoint; an overview of
the available key bindings can be displayed by pressing \texttt{"?"}.

When a slide includes a code snippet, the viewer will find it augmented
with semantic information (Figure \ref{fig:slides}). The syntax highlighting will reflect
the semantic meaning of tokens, just like in Isabelle/jEdit \cite{wenzel2014system} or Clide \cite{ring2014collaborative, lueth2013}. For
Isabelle theories, the proof states can be iterated (just like other
fragments) if the class \texttt{state-fragments} is added. It is also
possible to mark certain fractions of the code as fragments which will
behave like any other \textit{reveal.js} fragment but also affect the
semantic annotations. Errors are prominently rendered as a red
background behind the affected part of the snippet (Figure \ref{fig:error}), which is useful when
presenting a situation where an error is of importance (we found the usual
red underlining to be too innocuous). It is also
possible to view additional information about entities by selecting with the
mouse (or a finger when using a touch enabled device). The displayed
tool-tips are displayed in a way that is suitable for presentations, and
contain similar information to the tool tips in Isabelle/jEdit (Figure \ref{fig:hover}).
Since manually selecting can be cumbersome it is also possible to use
\emph{selection fragments}, which automate this by allowing to step through
predefined ranges as described in Section~\ref{sec:code}.

Furthermore, it is possible to alter the content of snippets just like in a
text editor. \footnote{%
Note that for security reasons, the Safari web browser does not allow keystrokes in
full screen mode, which unfortunately makes the browser unsuitable for presenting interactive
proofs.} When snippets are changed, the content is synchronised across all connected devices
as well as the underlying assistant, which then annotates the code with updated semantic
annotations, similar to the non-human collaborators in Clide. It is also possible that
someone from the audience alters the code with his own device, if the
speaker allows access to the presentation server\footnote{The default
  setting of serving to \texttt{localhost} prevents this.}. The changes will then be
synchronised to everybody else and become visible in the main presentation.
Snippets are synchronised continuously; \eg{} if there is one running code
example that is referenced across various slides, changes made in one slide might
affect the others. Overlapping snippets will always display consistent information.
If this is not desired, the snippets have to be separated into different origins.

\subsection{Publishing and distribution}

There are three intended ways of publishing slides. All three have
unique advantages and disadvantages. Thus, it is desirable to provide at least two
different ways to access slides for the audience.

\paragraph{Central presentation server:} The presentation can be provided on a central presentation
server which then also runs the Isabelle process. Viewers can access the presentation and play around with
proofs through their web browser. This approach requires a lot of resources on the presentation server
when many viewers access different presentations at the same time.

\paragraph{Local presentation servers:} It is possible to distribute a \textit{Cobra} presentation
as an archive file. This will require the viewers to install the
\textit{Cobra} command line tool as described above,
and install Isabelle locally.

\paragraph{PDF export:} Having active slides is not always desirable, especially
when it comes to just viewing and printing. In this case it is possible to export PDF. This can
be done the same way as in \textit{reveal.js}; \textit{Cobra} then takes care of the correct rendering of
snippets for printing.

\section{\textit{Cobra} internals}
\label{sec:int}

The PIDE framework \cite{wenzel2014system} which was developed together with the Isabelle/jEdit
integration has already been successfully integrated into the collaborative web
environment \textit{Clide} in previous work \cite{ring2014collaborative}. We were able to reuse
significant portions of the code to implement \textit{Cobra}. However, due to
the limitations at the time, while the \textit{Clide} server is
implemented in Scala, a strongly typed and reasonably well specified
language, the client was implemented in CoffeeScript, a thin wrapper language
compiling directly to JavaScript. Both CoffeScript and JavaScript are dynamically typed languages, with an object system based
on prototyping, and lacking a module system. While this makes these languages
suited for ``quick and dirty'' solutions, complex projects like \textit{Clide}
or \textit{Cobra} become hard to maintain and reason about. Recently, several independent
approaches, like TypeScript~\cite{typescript}, arose to remedy these shortcomings. A particularly interesting
solution in our context is \textit{ScalaJS} \cite{doeraene2013scala}, a JavaScript
back-end for the Scala compiler, which compiles Scala code to JavaScript
which can run in a web browser; it has officially been labelled
``ready for production'' by the creators of Scala in 2015.

By using \textit{ScalaJS} the \textit{Cobra} code base does not need to
include a single line of JavaScript. We do depend on JavaScript libraries
like CodeMirror (the editor component) or \textit{reveal.js} (the
underlying presentation framework), but were able to create Scala facade
types from existing TypeScript types. Especially the basis of
\textit{Clide}, the collaboration algorithm, is now identical on the JVM
and all connected browsers.

\paragraph{System architecture.}
\textit{Cobra} is designed as a client-server architecture. Unlike \textit{Clide}
which is based on the Play! framework, the \textit{Cobra} server is built
around akka-http~\cite{roestenburg2015akka}, a minimal HTTP library based on the Akka implementation of reactive streams~\cite{reactivestreams}.
This results in a reduced size and better speed of the application.
Upon connection, web clients load all static assets from the \textit{Cobra} server through plain HTTP, these include the compiled JavaScript of the client. After that, a WebSocket
connection is established which handles all further communication. The protocol is a
very fast and size-efficient binary protocol based on the boopickle library, which is itself
derived from the Scala Pickling project \cite{miller2013instant}.
This allows us to pass Scala defined algebraic data types around between client and server,
both of which are implemented in Scala. Since the PIDE framework is also implemented in Scala, the
integration is straightforward. The same holds for the Scala compiler. Haskell is
integrated by calling ghc-mod as an external command. To synchronise document states
across clients and assistants, the universal collaboration approach from
Clide is used (as described in \cite{ring2014collaborative}).

\section{Related Work}
\label{sec:rel-work}

There is a wealth of work on presenting formalised proof, going back right
to the start of the field. Early attempts were concerned with making proof
scripts more like mathematical texts rather than programming languages; one
of the earliest representatives of this was the Mizar prover. However, this
does not allow interaction with the proof script beyond it being checked by
the prover.
There have been various attempts at
presenting interactive views on proofs, most of which focus on specific
aspects, logics or systems:
the Jape prover~\cite{JAPE99a} allows direct manipulation of a natural
deduction proof visualised as a tree or box, Grundy and Back use
structured calculational reasoning \cite{Grundy1996,Back1997} where the
proof can be interactively explored along its hierarchical structure,
the Omega system visualised proofs interactively using proof trees
\cite{LouiLoui},
Theorema uses the computer algebra system Mathematica \cite{JFR4568},
and there have been various attempts to visualise geometric proofs using
diagrams \cite{wilson2005combining,Narboux2007,Ye2010}.

On a less conceptual and more technical level, there are systems which are
technologically similar to \textit{Cobra} in that they make use of web-based front-end technology:
ProofWeb was an early precursor \cite{Kaliszyk2007}, Clide offers a full
web interface for PIDE-enabled provers \cite{ring2014collaborative}, and
jsCoq moves the whole prover into the web browser by translating it into
JavaScript \cite{Arias2016}. However, while these are similar in some
regards, differences remain; \eg jsCoq is completely devoted to Coq, and
not generic like \textit{Cobra}.

Summing up, the following three characteristics distinguish \textit{Cobra} from
other systems: firstly, it focuses on presentations (slides) as
opposed to long-form text, which means that extraction of parts of the
proof plays a key role; secondly, it is interactive, allowing to change
content and display the results immediately; and thirdly, it is generic,
\ie{} usable with different provers or programming languages.

\section{Conclusion}
\label{sec:concl}

We have presented \textit{Cobra}, a framework which allows source code and
interactive proof scripts to be presented as active documents, with which
the viewer can interact. Our tool is ready for use in production; it is
distributed in binary form for all major operating systems at
\url{http://www.flatmap.net/cobra}\footnote{%
  Please note you need to have Java 8 installed to run the \textit{Cobra}
  binary.}, %
or alternatively in source code form at
\url{https://github.com/flatmap/cobra/}.

We envisage the following usage scenarios for \textit{Cobra}: Firstly, classroom teaching where a
prepared proof is presented to a group of students. Here, the advantages of
\textit{Cobra} are that the teacher can select those parts of the proof to
be presented, so the presentation remains compact and convenient to follow,
avoiding cognitive overload with lots of unnecessary details. The teacher
can further build up the proof gradually, like on a blackboard, but with
the safety net of the Isabelle proof checker in the background. Secondly,
teaching in small groups where the teacher develops a proof together with a
group of students. Here, the proof can be developed collaboratively, with
everybody contributing while the presentation serves as the focus of
attention for all participants. Thirdly, self-study when readers (students,
fellow researchers, or reviewers) can interact with the presentation,
exploring the effects of changes. Finally, research talks at a workshop
such as UITP, where researchers can present their work with greater
confidence, and can pick up questions from the audience by demonstrating
effects of changes as they might be suggested from the floor.
In all of these scenarios, \textit{Cobra} brings added value over the
current state of the art, where interactive proofs are presented as passive
documents.

In \emph{future work}, we plan to include other proof assistants. The canonical way to connect other proof assistants is via the
PIDE framework. As a PIDE integration exists \cite{tankink2014pide}, it should be feasible to
integrate Coq. In addition, it would be beneficial to reduce the overhead further by introducing a
simplified syntax to describe slides, possibly based on Markdown. Further,
we plan to explore if more inter-dependencies between the snippets and the presentation itself could be
allowed. This would allow for even richer presentations, where the structure of a presentation depends on
the contained proofs or generated graphics visualise dynamic aspects of a proof.

\bibliographystyle{eptcs}
\bibliography{references,more}

\end{document}